\documentclass[prb,twocolumn,superscriptaddress,floatfix,showpacs]{revtex4-1}
\usepackage{graphicx,amsfonts,amssymb,amsmath, hyperref}

\newif\ifhyper
\hypertrue
\ifhyper
\hypersetup{
   citecolor = {green},
   colorlinks = {true}, 
   urlcolor = {blue} 
}
\fi

\newcommand{\beq}{\begin{equation}}
\newcommand{\eeq}{\end{equation}}
\newcommand{\beqa}{\begin{eqnarray}}
\newcommand{\eeqa}{\end{eqnarray}}

\def\Longarrow{\protect\@lra}
\def\@lra{\relbar\joinrel\relbar\joinrel\relbar\joinrel%
          \relbar\joinrel\rightarrow}

\begin{document}

\title{Density of Yang-Lee zeros in the thermodynamic limit from tensor network methods}

\author{Artur Garc\'ia-Saez}
 \affiliation{C. N. Yang Institute for Theoretical Physics and Department of Physics and Astronomy,
State University of New York at Stony Brook, NY 11794-3840, USA}

\author{Tzu-Chieh Wei}
 \affiliation{C. N. Yang Institute for Theoretical Physics and Department of Physics and Astronomy,
State University of New York at Stony Brook, NY 11794-3840, USA}

\begin{abstract}
 The distribution of Yang-Lee zeros in the ferromagnetic Ising model in both two and three dimensions is studied on the complex field plane directly in the thermodynamic limit via the tensor network methods.
 The partition function is represented as a contraction of a tensor network 
and is efficiently evaluated with an iterative tensor renormalization scheme. The free-energy density and the magnetization are computed on the complex field plane.
Via the discontinuity of the magnetization, the density of the Yang-Lee zeros is obtained to lie on the unit circle, consistent with the Lee-Yang circle theorem. Distinct features are observed at different temperatures---below, above and at the critical temperature. Application of the tensor-network approach is also made to the $q$-state Potts models in both two and three dimensions and a previous debate on whether, in the {\it thermodynamic\/} limit, the Yang-Lee zeros lie on a unit circle for $q>2$ is resolved: they clearly do not lie on a unit circle except at the zero temperature. For the Potts models ($q=3,4,5,6$) investigated in two dimensions, as the temperature is lowered the radius of the zeros  at a fixed angle from the real axis shrinks exponentially towards unity with the inverse temperature. 
\end{abstract}

\maketitle

\section{Introduction}
More than half a century ago Yang and Lee proposed a new approach for studying phase transitions in a gas by examining the zeros of the grand partition function in the complex fugacity plane~\cite{YangLee1}.   In the thermodynamic limit,
these zeros, which shall be referred to as the Yang-Lee zeros, may lie arbitrarily 
close to certain points on the real axis, marking where phase transitions appear.
Inside a region clear of zeros, no phase transitions can occur.
Thus, the study of the location of the zeros in the complex plane determines the transition points in the real axis~\cite{mccoy}.
In a subsequent paper~\cite{YangLee2}, Lee and Yang showed, among other things, that the zeros of the partition for the ferromagnetic Ising model lie on a unit circle of the complex field (or more precisely the complex $z$-plane, where $z\equiv \exp(-2\beta h)$ with $\beta=1/{\rm k_B} T$ and $h$ the external field), which is referred to as the Lee-Yang theorem.  The equation of the state can be obtained via the knowledge of the distribution of the Yang-Lee zeros on the unit circle. At very high temperatures, the zeros do not cover the whole circle, but only the segment of the arc around the angle $\theta=\pi$. As the temperature is lowered to the transition temperature $T_c$, the zeros move and pinch the real axis at $\theta=0$, the field value (in this case zero) of which corresponds to the phase transition point. 

The study of partition function zeros, as well as generalization of the circle theorem, has been extended to higher spins and other models (whose list is hard to exhaust here)~\cite{abe,Asano,AsanoPRL,Suzuki,Griffiths69,HeilmannLieb,SuzukiFisher,fisher78,IzyksonPearsonZuber,Car85,MittagStephen,WoodTurnbullBall,glumac,GlumacUzelac,KC98,Assis} and has provided useful insights to phase transitions, even in QCD~\cite{QCD}. The consideration of zeros in the complex temperature plane was initiated by Fisher~\cite{FisherZ} and these zeros are called Fisher zeros~\cite{mccoy,FST98,matveev}.
Properties of the distributions of zeros also give characterization of first-order phase transitions~\cite{Biskup} as well as higher-order phase transitions and scaling relations~\cite{janke}.  The Lee-Yang theorem has its impact also beyond statistical physics; it has incited mathematical theory connected with the Laguerre-P\'olya-Schur theory of linear operators, possible connection to the Riemann hypothesis on the zeta functions~\cite{Newman,Knauf}, and computational complexity of computing averages~\cite{sinclair}, etc.
 Even though the Yang-Lee zeros lie on the complex plane, their density was inferred from the magnetization data in one experiment~\cite{Binek98}, and very recently it is found that the individual zeros of classical Ising models can be detected by using a quantum spin as a probe that couples to the classical spins~\cite{wei12,peng}. 

Yang-Lee zeros can be solved exactly on small system sizes and the thermodynamic limit is inferred from extrapolation. For most models, however, there is no analytic expression for the distribution of the zeros and to locate these zeros accurately in the thermodynamic limit remains a challenge.   
It was known that computing averages such as the magnetization exactly for even the ferromagnetic Ising model on the complex plane is \#P hard~\cite{sinclair}.
Here we introduce the 
tensor network (TN) techniques for computing the density of the Yang-Lee zeros in the complex field plane directly in the thermodynamic limit. 
The density is proportional to the discontinuity of the magnetization on the complex plane~\cite{YangLee2}, and the magnetization is computed with controlled approximations. Tensor network methods, including the density matrix renormalization group (DMRG), matrix product states (MPS), and other tensor product or tensor network states, have become very useful numerical tools in both classical and quantum spin systems~\cite{dmrg,nish1,nish2,VC04,vidal07,tn}. In essence, physical observables are expressed as contraction of a tensor network. However, the contraction in two and higher dimensions is a computationally hard problem, but it can be approximated and the error can be controlled by reducing truncation, such as during the real-space coarse-graining or renormalization group (RG) procedure over the tensors describing the system~\cite{levin,wen,Orus12,xiangPRB,gsl,xiangpotts1}.  We note that the particular partition function zero closest to the positive real axis
has been explored with TN for the one-dimensional Schwinger model on a finite lattice~\cite{SK14}. But  for our purpose we use the method appropriate for the infinite system to directly probe the density of zeros in the thermodynamic limit.

In particular, we study ferromagnetic Ising and $q$-state Potts models~\cite{wu} in both two and three dimensions. We observe clearly that the magnetization has a dicontinuity at the unit circle, consistent with the Lee-Yang circle theorem for the Ising model. The distribution of the zeros on the unit circle shows distinct features at different temperature regimes: $T>T_c$, $T=T_c$ and $T<T_c$.  At $T>T_c$ the zeros occupy part of the circle and move along it towards the real axis as $T$ decreases. They pinch $z=1$ at $T=T_c$ and the difference with $T<T_c$ is manifest in the distribution of the zeros. 
From the density we obtain good estimates for the magnetization-field exponent $\delta$ for both two and three dimensions. 
 For the $q$-state Potts models in two and three dimensions our results  demonstrate that the zeros clearly do not lie on a unit circle in the thermodynamic limit for $q>2$ and $T>0$, in agreement with Kim and Creswick~\cite{KC98}, whose results were questioned previously due to the small system sizes~\cite{monroe}. Our results also show that the deviation from the unit circle, in terms of the radius at a fixed angle, is decaying exponentially with the inverse temperature $\beta$, consistent with the fact that the location of the zeros shrinks to the unit circle at zero temperature.

The remaining organization of the paper is as follows. In Sec.~\ref{sec:zeros}, we give a slightly more expanded but elementary introduction to the Yang-Lee zeros, and the readers can skip it if they are already familiar with these ideas. In Sec.~\ref{sec:free} we review the RG algorithm  that we employ in this paper,  with a detailed exposition of the 
thermodynamic limit calculations of the free energy density. A numerical RG process is applied 
to the Ising model in both a square and cubic lattices, with computation of the magnetization described in Sec.~\ref{sec:density},
to show the Yang-Lee zeros on the complex plane. From the magnetization
we directly obtain the density of zeros along the unit circle. In particular the distribution of the zeros at $T=T_c$ is used to estimate the magnetization-field exponent $\delta$. For $T>T_c$ we study in Sec.~\ref{sec:edge}
the `edge singularity' of the density of zeros. In Sec.~\ref{sec:potts} 
we extend our results to the Potts models, and we study the location of zeros in 2D and 3D lattices. We conclude in Sec.~\ref{sec:conclusions}.

\section{Yang-Lee zeros: an elementary introduction}
\label{sec:zeros}
Here we give an elementary introduction to the Yang-Lee zeros. Readers who are familiar with Yang-Lee zeros can skip this section. 
If we consider  a system of $N$ Ising spins with the Hamiltonian, 
\begin{equation}
H=-\sum_{\langle i,j\rangle} s_i s_j -h\sum_i s_i,
\end{equation}
where $s_i=\pm 1$ is a classical variable and
$h$ is  an external field. The partition function at a temperature $T=1/\beta$ (where we have set $k_B=1$) is
\begin{equation}
Z(\beta,h)={\rm Tr}(e^{-\beta H})=e^{ N \beta h} \sum_{n=0}^N P_n z^n,
\end{equation}
where in the second equality we have re-arranged the contribution to $Z$ in terms of the number $n$ of down spins and their associated value $P_n$ at zero field and we have defined $z\equiv \exp(-2\beta h)$. Since $P_n$ is independent of $h$ (and is real and positive), one can factorize the polynomial ${\cal P}(z)\equiv \sum_{n=0}^N P_n z^n=c_0 \prod_{n=1}^N (z-z_n)$, where $z_n$'s are the zeros of ${\cal P}(z)$, which are referred to as the Yang-Lee zeros and $c_0$ is a positive constant. 

Lee and Yang proved that for any ferromagnetic Ising model
the zeros $z_n$ lie on a unit circle, namely $z_n=e^{i\theta_n}$. This is referred to as the Lee-Yang circle theorem~\cite{YangLee2}. The consideration of the zeros and the Lee-Yang theorem was generalized to other models and higher spins~\cite{abe,Asano,AsanoPRL,Suzuki,Griffiths69,HeilmannLieb,SuzukiFisher,fisher78,IzyksonPearsonZuber,Car85,MittagStephen,WoodTurnbullBall,glumac,GlumacUzelac,KC98,Assis}. 

How are the Yang-Lee zeros related to phase transitions? The free-energy density is 
\begin{eqnarray}
f(\beta,h)&=&-\ln Z(\beta,h)/(N\beta)\\
&=&-h -\frac{\ln c_0}{N\beta} -\frac{1}{N\beta}\sum_{n=1}^N \ln(z-z_n).\end{eqnarray}
In a region  free of zeros, the free energy is analytic and therefore there cannot be any singularity, and hence on the real axis contained in this zero-free region, no phase transitions occur. The Yang-Lee zeros close to the positive real axis of $z$ signal the location of the phase transitions. From the free energy, one can obtain the magnetization (which in general has a complex value for complex $z$),
\begin{equation}
m(\beta,h)=-\frac{\partial f}{\partial h}=1-\frac{2z}{N} \sum_{n=1}^N \frac{1}{z-e^{i\theta_n}}.
\end{equation}
In the limit $N\rightarrow\infty$, the zeros form a continuum on the unit circle, with the density of zero per angle denoted by $g(\theta)$. Since any complex zero $z_n$ has a corresponding complex conjugate partner $\overline{z_n}$, the density of zero has the symmetry that $g(-\theta)=g(\theta)$, and, employing this, one can re-rewrite the free-energy density as follows,
\begin{eqnarray}
f(\beta,h)=-h -\frac{1}{\beta}\int_{0}^\pi d\, \theta \,g(\theta)\ln({z^2-2z\cos\theta +1}),\end{eqnarray}
and the magnetization as follows
 \begin{equation}
m(\beta,h)=1-{4z} \int_{0}^\pi d\theta \,g(\theta)\frac{z-\cos\theta}{z^2-2z\cos\theta +1},
\end{equation}
which we have chosen the normalization that
\begin{equation}
\int_{0}^{2\pi} d\theta\, g(\theta)=1.
\end{equation}
The thermodynamics of the system therefore can be deduced if the density of zeros $g(\theta)$ is known on the unit circle.  We remark that, however, except in one dimension, the analytic expression of $g(\theta)$ is generally not known. But the Yang-Lee picture for phase transitions is very visual and intuitive in terms of partition function zeros; see e.g. Figs.~\ref{fig:free_energy},~\ref{fig:mag_2D}\&~\ref{fig:mag_3D}. Approximation schemes such as  series expansion have been applied to the density~\cite{KG71}. 
Using an electrostatic analogy, Lee and Yang showed that the density of zeros is related to the discontinuity of the magnetization across the unit circle, i.e.,
\begin{equation}
g(\theta)=-\frac{1}{4\pi} {\rm Re}\left(\lim_{r\rightarrow 1+}m(z=r\,e^{i\theta}) -\lim_{r\rightarrow 1-}m(z=r\,e^{i\theta}) \right).
\end{equation}  
Since the free-energy density is expressed in terms of the zero density, other thermodynamic quanities in addition to the magnetization and the susceptibility, such as the specific heat and the entropy density can also be obtained once the density is known. Our numerical calculation, to be discussed below, is based on the above relation between the magnetization and the density and is to obtain the latter via evaluation of the magnetization directly on the complex plane with the tensor network methods. We remark that we shall loosely refer to the real part of the magnetization as the magnetization when no confusion arises, as the imaginary part of the magnetization is never needed in our discussions.

More than a decade after the results by Lee and Yang~\cite{YangLee2}, Fisher initiated the study of the zeros defined on the complex temperature plane, i.e., the so-called Fisher zeros~\cite{FisherZ}.
Analyzing the behavior of Yang-Lee and Fisher 
zeros also allows  characterization of first-order phase transitions~\cite{Biskup} as well as second and higher-order phase transitions and scaling relations~\cite{janke}. The interest of Lee-Yang-like theorem for partition function zeros goes beyond statistical physics, such as in mathematics~\cite{Newman,Knauf} and computational complexity~\cite{sinclair}. But further detailed explanation on these will take us awry from the main purpose of this work and the readers are referred to the cited references and more references therein for further discussions. This section serves as a brief and elementary introduction to the Lee-Yang zeros that we shall explore in later sections.

\section{Free energy from Tensor Renormalization}\label{sec:free}

The partition function $Z={\rm Tr}\,\exp\{-\beta H\}$ of a classical system with local interactions can be expressed as a contraction of a tensor 
network of low rank and of low dimensions. The expectation $\langle O\rangle$ of local observables $O$, $\langle O\rangle={\rm Tr}\, (O \, \exp\{-\beta H\})/Z$, is then a ratio of contractions of two TNs. 
This efficient representation serves as the starting point for a coarse-graining 
process for computing physical observables from the tensor. 
The contraction of a TN is in general a hard problem, and numerical approximations are in order~\cite{sharp}.
In recent years a number of schemes have been proposed to approximate $Z$ in an efficient manner; see e.g. Refs.~\cite{tn,OrusReview} and references therein. 
The RG process consists of a decimation of the lattice at each step and such decimation process is iterated, which gives rise to efficient and accurate 
predictions of thermodynamic quantities,  even close to the phase transitions.

\begin{figure}
\includegraphics[width=7cm]{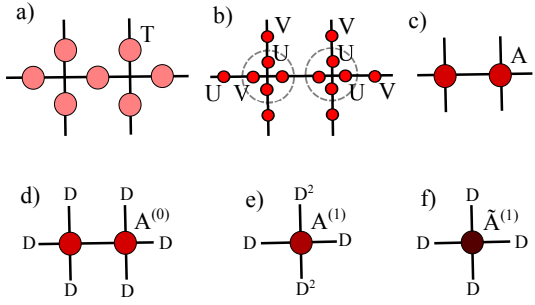}
\caption{(Color online) The free energy of a classical lattice system is expressed as a Tensor Network,
and evaluated using a RG iterative process. (a) The Tensor $T$ of rank 2 is obtained from the Hamiltonian (see Eq.~\ref{eq:tensor} in the text). By means of a decomposition (b), tensors $U$ and $V$ are obtained, and combined (c) to form tensor $A$.
An RG step over the square lattice consists in the coarse grain of a set of tensors into 
a single site tensor. {\emph I.e.} combining two tensors horizontally (d), a new tensor $A$ is created (e). After truncation, the new tensor (f) is 
used again in an iterative process.} \label{fig:diag}
\end{figure}

\begin{figure}
\includegraphics[width=4cm]{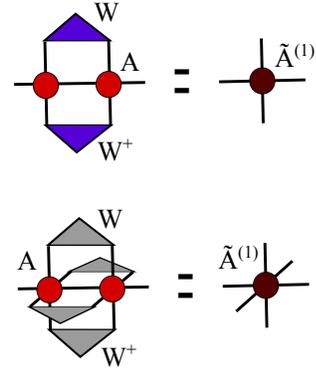}
\caption{(Color online) The HOTRG algorithm proceeds at each step combining 2 tensors along a preferred direction, 
and extracting unitaries on the combined indices. These unitaries $W$ are truncated and inserted in order 
to reduce the dimension of the auxiliary indices. This prevents an explosion of the number of parameters 
required, while keeping accurate approximations to $Z$. Similar operations are required either for 
2D lattices (top, with tensors of rank 4) and for 3D lattices (bottom, tensors of rank 6).} \label{fig:hotrg}
\end{figure}

Let us take for example the Ising model with nearest-neighbor interaction and a local field,
\beq
H = \sum H_{i,j} = \sum_{\langle i,j\rangle}[-s_i s_j - \frac{h}{n_b}  (s_i+s_j)]
\eeq
where $\langle i,j\rangle$ represents the nearest-neighbor pair $i$ and $j$, $n_b$ is the number of neighbors, $n_b=2d$ for square ($d=2$) and cubic ($d=3$) lattices.
Its partition function $Z$ can be expressed as a tensor network; the explicit description of the tensors 
forming the network are obtained straightforwardly from the Hamiltonian, 
\beq
Z = {\rm Tr}\,\exp (-\beta H) ={\rm Tr}\, \prod_{\langle i,j\rangle} \exp\{-\beta H_{i,j}\} 
\eeq
can be decomposed as the tensor trace of
the product of identical tensors $T$,
\beq
Z = {\rm tTr} \prod TT\ldots T,
\eeq
where {\rm tTr} denotes the tensor trace, i.e.,  summing over all degrees of freedom, and each tensor $T$ (lying on each of the edges of, e.g., the square lattice, see Fig.~\ref{fig:diag})
has the expression
\beq
T = \left( \begin{array}{cc}
\exp\{\beta + h/d\} & \exp\{-\beta\} \\
\exp\{-\beta\} & \exp\{\beta - h/d\} \end{array}\right) \label{eq:tensor}
\eeq
The product operation transforms the tensor network into a number by contracting i.e. summing over
all the corresponding degrees of freedom of neighboring sites.
Each tensor $T$ has rank 2 and is located on an edge of the square lattice.
For convenience of an RG iteration that directly preserves the square and cubic lattice structure of the tensors, we shall instead construct new tensors $A$ that are located on each site and the trace of them is to sum over
degrees of freedom on edges. To do this, we first use a simple decomposition (such as singular-value decomposition) of each tensor $T$ as 
\beq
T_{i,j} = \sum_\mu U_{i,\mu}V_{j,\mu},
\eeq
and then we form on each vertex  a new tensor $A$ from the contraction of four tensors (chosen from $U$ or $V$) in the square lattice 
\beq
A_{u,d,l,r} = \sum_i U_{i,u}V_{i,d}V_{i,l}U_{i,r}.
\eeq
(For the simple cubic lattice, $A$ will be a rank-6 tensor obtained from six $U$ or $V$ tensors.)
Therefore,  the partition function of the Ising model on a square lattice involves a single tensor 
$A$ repeated in the infinite lattice, and is an exact representation of the partition function of the system, which
is then evaluated via an RG iterative process to be described below.
The basic structure of an RG process to compute $Z$ is a partition of the lattice into small plaquettes.
At each step, plaquettes are coarse-grained to obtain an effective lattice of smaller size. In the limit of many
iterations of this process, plaquettes become invariant upon further renormalization step. At this effective 
thermodynamic limit we can easily evaluate the free energy per site or local observables using a single plaquette.
While the general picture of RG methods is essentially equivalent but differs in the specific algorithmic implementations, in this paper 
we employ the higher-order tensor RG (HOTRG) \cite{hotrg},  which among other schemes has shown accurate evaluation of physical observables. 
The basic ingredient of the algorithm is presented in Fig.~\ref{fig:hotrg}, where a pair of tensors $A$ is coarse-grained into a single tensor $\tilde{A}$. In practice, the coarse-graining procedure will be done iteratively for the horizontal direction followed by the vertical direction in the 2D square lattice, and similarly for the three directions in the 3D simple cubic lattice.

From the RG flow one directly obtains approximations to the free energy per site~\cite{HN}
\beq
f = \frac{-1}{N\beta}\log{Z}.
\eeq
Before starting the RG, we have a TN representing the Ising model in a square lattice 
with initial equal tensors $A^{(0)}$; as we perform the RG contraction numerically, we keep 
the tensors under machine representation factorizing $A^{(0)}$ at each step as follows,
\beq
\label{eqn:factor}
A^{(0)}=|\alpha_0|\tilde{A}^{(0)}.
\eeq
From this the RG process effectively shrinks the lattice sites by half and produces a sequence of new tensors $A^{(1)}, A^{(2)}...$, and
each of these tensors is also factorized accordingly to bound the numerical values similar to the above Eq.~(\ref{eqn:factor}). Let us define 
\beq
G^{(0)}\equiv\ln|\alpha_0|
\eeq
and the corresponding $G^{(n)}$. The partition function in the TN picture then reads
\beq
Z\equiv[A^{(0)}]^N = e^{N G^{(0)}}[\tilde{A}^{(0)}]^N,
\eeq
where we have defined the notation $[A]^N$ to indicate a contraction of tensor network described by $A$ over $N$ sites.
By applying an RG step over $\tilde{A}^{(0)}$ we have
\beq
Z\equiv e^{N G^{(0)}}[A^{(1)}]^\frac{N}{2} = e^{N G^{(0)}} e^{\frac{N}{2} G^{(1)}} [\tilde{A}^{(1)}]^\frac{N}{2}
\eeq
where $\tilde{A}^{(1)}$, after a single RG step, spans a smaller lattice or more precisely half the lattice with $N/2$ sites. 
This process can be iterated and the free energy per site can be written as a function of all the prefactors $G^{(k)}$ as follows,
\beq
-\beta\,
f = \sum_{k=0}^n \frac{G^{(k)}}{2^k} + \frac{1}{N}\ln\{[\tilde{A}^{(n)}]^{N/2^n}\},
\eeq
where $n$ the total number of RG steps that has been carried, and the second term vanishes exponentially for large $n$.  Notice that without truncation the dimension of each index of the tensor $\tilde{A}$ will increase exponentially with $n$, and the common practice is to limit the maximal dimension to some number denoted by $D_{\rm cut}$~\cite{xiangPRB}. 
One can increase $D_{\rm cut}$ to see whether the computed observables converge.
By the above procedure, we thus obtain the free energy per site (at any complex field $h$ value) solely from the evaluation of the prefactors 
$G^{(k)}$ along the RG flow. In our calculations, the estimation of the free energy 
converges after only a few RG steps (typically $n\sim20$). We note that previous application of the tensor network methods for $f$ has been focused on real $h$ values and high precision for observables has been obtained~\cite{nish1,nish2}.

\begin{figure}
\includegraphics[width=7cm]{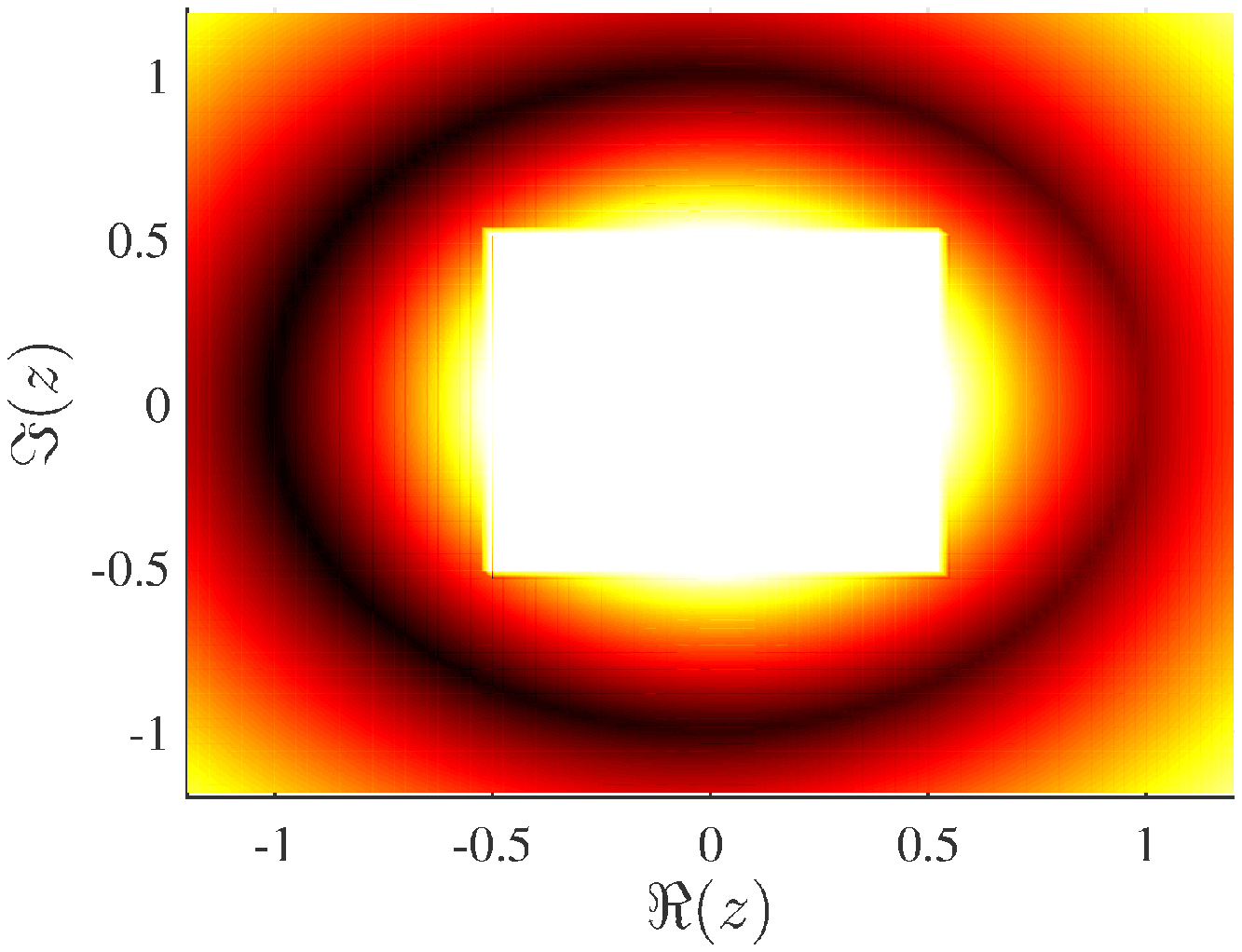}
\includegraphics[width=7cm]{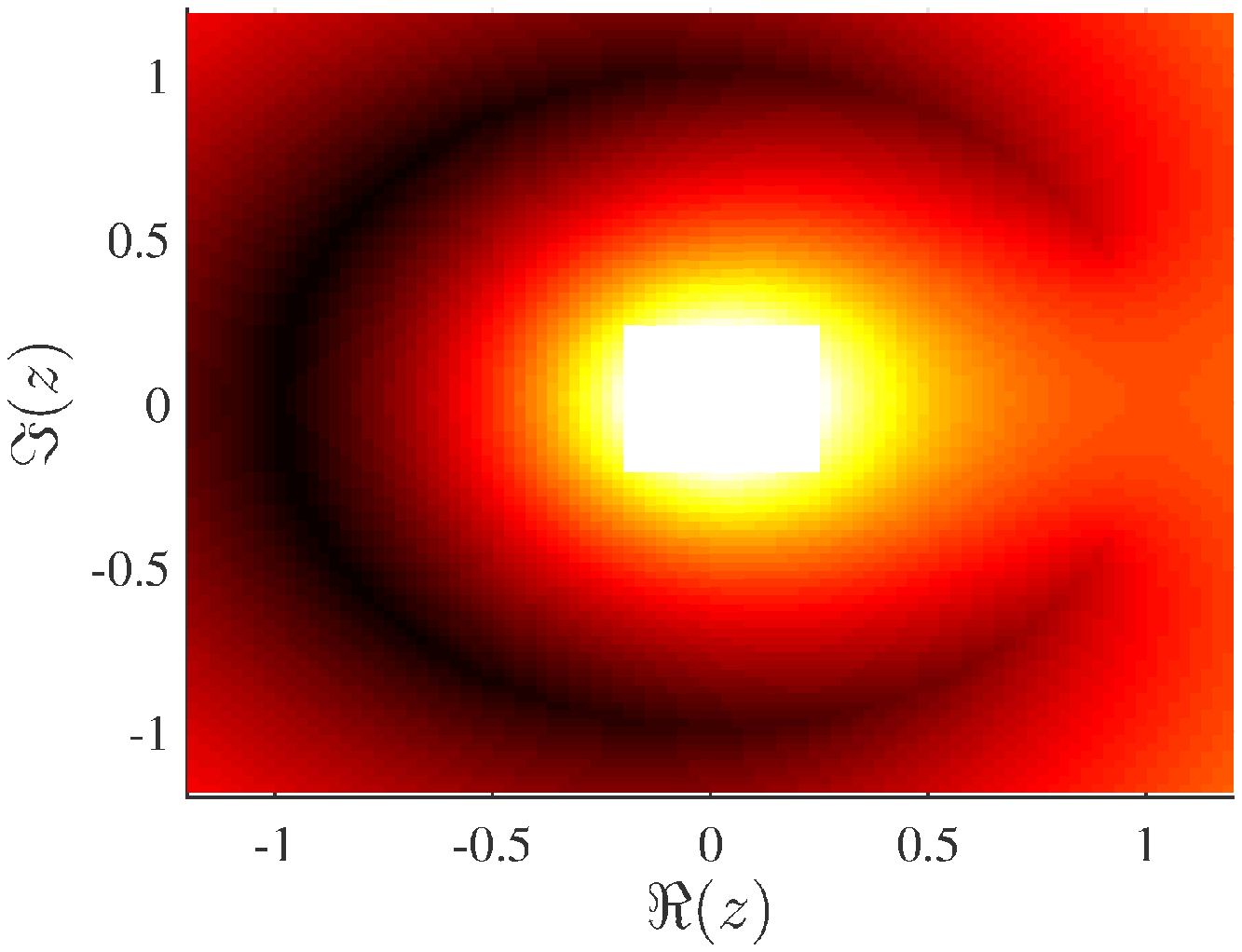}
\caption{(Color online) Free energy per site $f$ in the complex plane defined by $z=\exp{(-2h\beta)}$ for $\beta=\beta_c$ (top)
and $\beta=\beta_c/2$ (bottom), as obtained from Tensor Network calculations with $D_{cut}=10$. 
Darker regions around the unit circle correspond to minimum values of $f$. 
For $\beta<\beta_c$ this minimum does not contact the real axis at $z=1$. The central region possesses very large values of free-energy density and is removed so to enhance contrast around the unit circle.}\label{fig:free_energy}
\end{figure}

Would the TN methods work for complex field values? We use the above method to perform a calculation of the free energy per site $f$
on the {\it complex\/} plane defined by $z=\exp\{-2\beta h\}$. 
The result is shown in Fig.~\ref{fig:free_energy} for the Ising model in a 2D square lattice, for illustration, at two different temperatures. 
For any temperature, one always observes the minimum of $f$ at exactly the unit circle, a verification of the Lee-Yang theorem.
For $T<T_c$, the minimum of $f$ is located exactly along the unit circle, with a uniform density for any value of $\theta$ with $z\equiv r\,\exp(i\,\theta)$ at $r=1$. 
By increasing $T$ above $T_c=2/ \log(1+\sqrt{2})$, the values of $f$ around the positive real axis start to increase, 
and no discontinuity is observed  near $\theta=0$ at the unit circle; observe that the dark region recedes from the real axis close to unity. From the point of view of the magnetization (to be discussed in the next section), 
this translates into the disappearance of  phase transition at this high temperature.
These results demonstrate  how tensor network methods can be useful in the study of
partition function zeros, via the free
energy $f$ on the complex plane, obtained after an efficient RG contraction process. In the next section we shall show that the computation of magnetization leads to a more precise probe of the location of zeros and their density. 

\section{Magnetization and Density of Yang-Lee zeros in the Ising model}\label{sec:density}

\begin{figure}[ht]
\includegraphics[width=8.5cm]{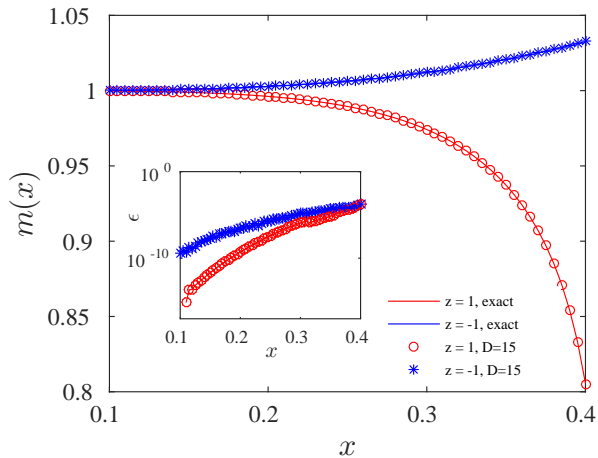}
\caption{(Color online) Magnetization as a function of $x\equiv\exp{(-2\beta)}$ at the points $z=1$ and $z=-1$ for the 2D Ising model. Solid lines are obtained from the analytical expressions in Eqs.~(\ref{eq:z1}) and~(\ref{eq:z-1}), and marks are obtained using Tensor Networks with $D_{cut}=15$. While the precision decreases around the transition point, 
it remains below $10^{-4}$ for this choice of $D_{cut}$ (see inset). }\label{fig:mag_exact}
\end{figure}

While one can obtain physical quantities directly from the derivatives of the free energy, 
the RG algorithms also provide a direct approach to compute expectation values of local observables such as the energy and magnetization.
 In order to compute the magnetization we define a new tensor $B$
\beq
B_{u,d,l,r} = \sum_i m_i U_{i,u}V_{i,d}V_{i,l}U_{i,r}\label{eq:mag_tensor}
\eeq
where $m_1=+1$ and $m_{-1}=-1$, accounting for the local magnetization on a single site.
Thus, calculations of the single site magnetization imply evaluation of the expression
\beq
m = \frac{1}{Z} {\rm Tr} (s_i e^{-\beta H})\equiv \frac{1}{Z}\langle M\rangle,\label{eq:m}
\eeq
involving two contractions of a TN: one contraction computes the norm $Z$ (using exactly the tensors 
defined for the calculation of the free energy), while the second contraction differs from the first one by only the single site tensor defined 
in Eq.~(\ref{eq:mag_tensor}) at only one site. We note that due to the complex field, the magnetization $m$ is not necessary real and nor restricted in the range $[-1,1]$.  Along the RG process, $\langle M\rangle$ is balanced by a prefactor $|\alpha_k|$, obtained from the RG
flow  from $Z$, and that keeps the numerical process bounded at each step. We use the ratio Eq.~(\ref{eq:m}) along the RG procedure
to ensure convergence, which normally takes about $50$ steps.
Performing two contractions after the RG has converged, we obtain a direct measurement of the magnetization in the state 
represented by the TN. This allows a complete study of the magnetization for
all complex values of the magnetic field $h$.

\begin{figure}[ht]
\includegraphics[width=8.5cm]{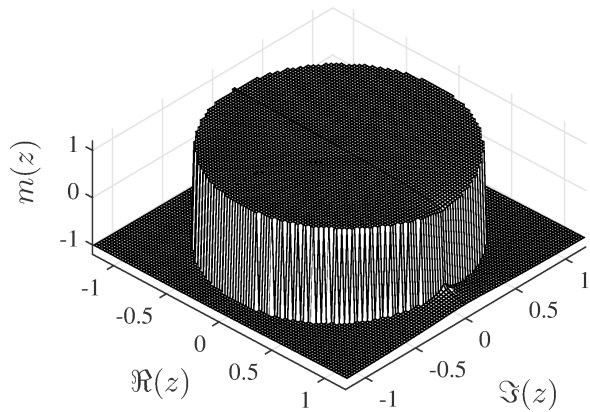}\\
\includegraphics[width=8.5cm]{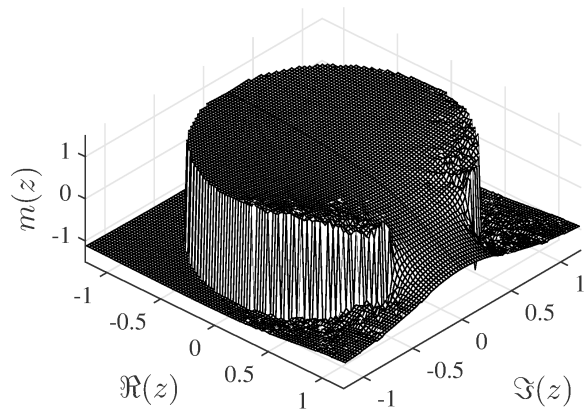}
\caption{Magnetization per site in the complex plane defined by $z=\exp\{-2\beta h\}$, as
obtained from the TRG with $D_{cut}=8$,  for the Ising model on the 2D  
square lattice. The temperature is fixed at $\beta=\beta_c$ (top) and 
$\beta=\beta_c/2$ (bottom). A discontinuity
in $m$ appears only at points on top the unit circle, and for a range of angles
$\theta$ that changes with the temperature.}\label{fig:mag_2D}
\end{figure}

For the Ising model on the 2D square lattice, only two exact solutions of $m$ are known in the complex plane~\cite{yang}: 
(i)~the seminal solution at $z=1$,
\beq
m(z=1) = \left[ \frac{1+x^2}{(1-x^2)^2}(1-6x^2+x^4)^\frac{1}{2})\right]^\frac{1}{4} \label{eq:z1}
\eeq
and (ii)~on the opposite side of the unit circle, at $z=-1$,
\beq
m(z=-1) = \left[ \frac{(1+x^2)^2}{1-x^2}(1+6x^2+x^4)^{-\frac{1}{2}})\right]^\frac{1}{4} \label{eq:z-1}
\eeq
with $x\equiv\exp{(-2\beta)}$.
We use these exact analytic results to benchmark the performance of the RG contraction.
In Fig.~\ref{fig:mag_exact} we present calculations of $m$ vs. $x$ at points $z=1$ and $-1$, and from comparison with the exact results, we find that 
the error of $m$ (as evaluated for different $D_{cut}$) is bounded below
$10^{-4}$, even close to the transition point. This clearly demonstrates that TN methods can
provide an accurate picture of the magnetization for complex values of the magnetic field. 
We remark that close to a phase transition, numerical accuracy can be improved by increasing the bond dimension ($D_{\rm cut}$).  Furthermore, more sophisticated RG algorithms can be employed~\cite{srg,hotrg}. Eqs.~(\ref{eq:z1}) and~(\ref{eq:z-1}) already provide useful information regarding the properties of the model and the partition function zeros.
 According to Eq.~(\ref{eq:z1}), at $z=1$ (or equivalently $h=0$) there is a critical value of $x$ (or equivalently $k_B T_c\equiv 1/\beta_c$) beyond which no magnetization is present, showing a phase transition. 
This transition for the 2D Ising model appears at $\beta_c=\ln{(1+\sqrt{2})/2}$. At $\theta=\pi$, however,
$m$ always increases with the temperature (with a value becoming larger than unity increasingly), and a divergence builds up here in the 
local magnetization at very large temperature $T$.  This buildup of divergence also implies the buildup of partition function zeros; see Eq.~(\ref{eq:density}) and  further discussions below.

In Fig.~\ref{fig:mag_2D} we plot
for the 2D Ising model  the magnetization in the complex $z$ plane. It 
shows a discontinuity only at exactly the unit circle. For the magnetization at two different temperatures we can 
observe how the temperature affects this discontinuity. For $T<T_c$ this discontinuity appears all along  the unit circle, 
even close to the positive real axis. However, at a higher temperature $T>T_c$, the discontinuity around $\theta=0$ vanishes,
and a smooth region emerges around there close to the positive real axis.

\begin{figure}[ht]
\includegraphics[width=8.5cm]{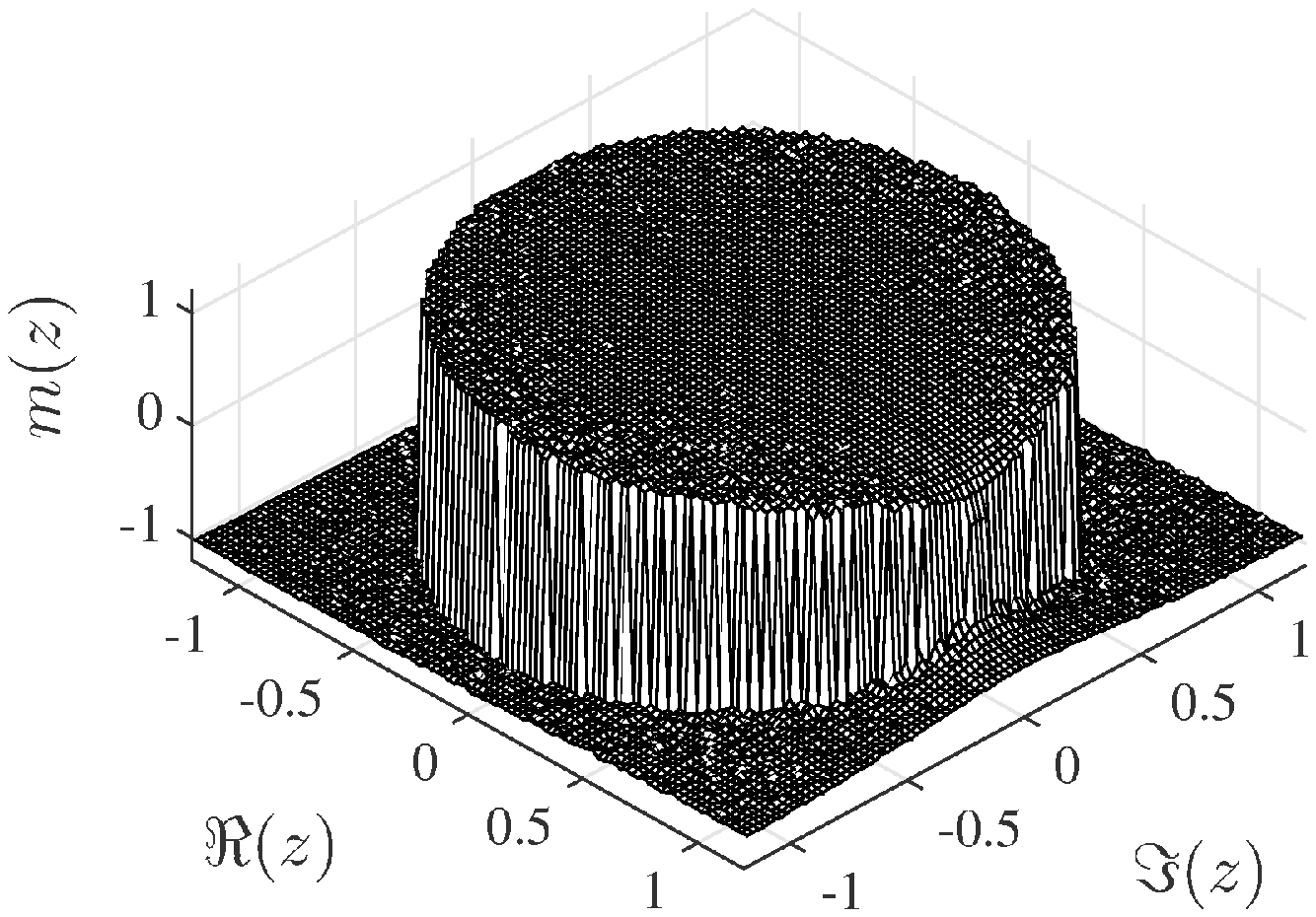}\\
\includegraphics[width=8.5cm]{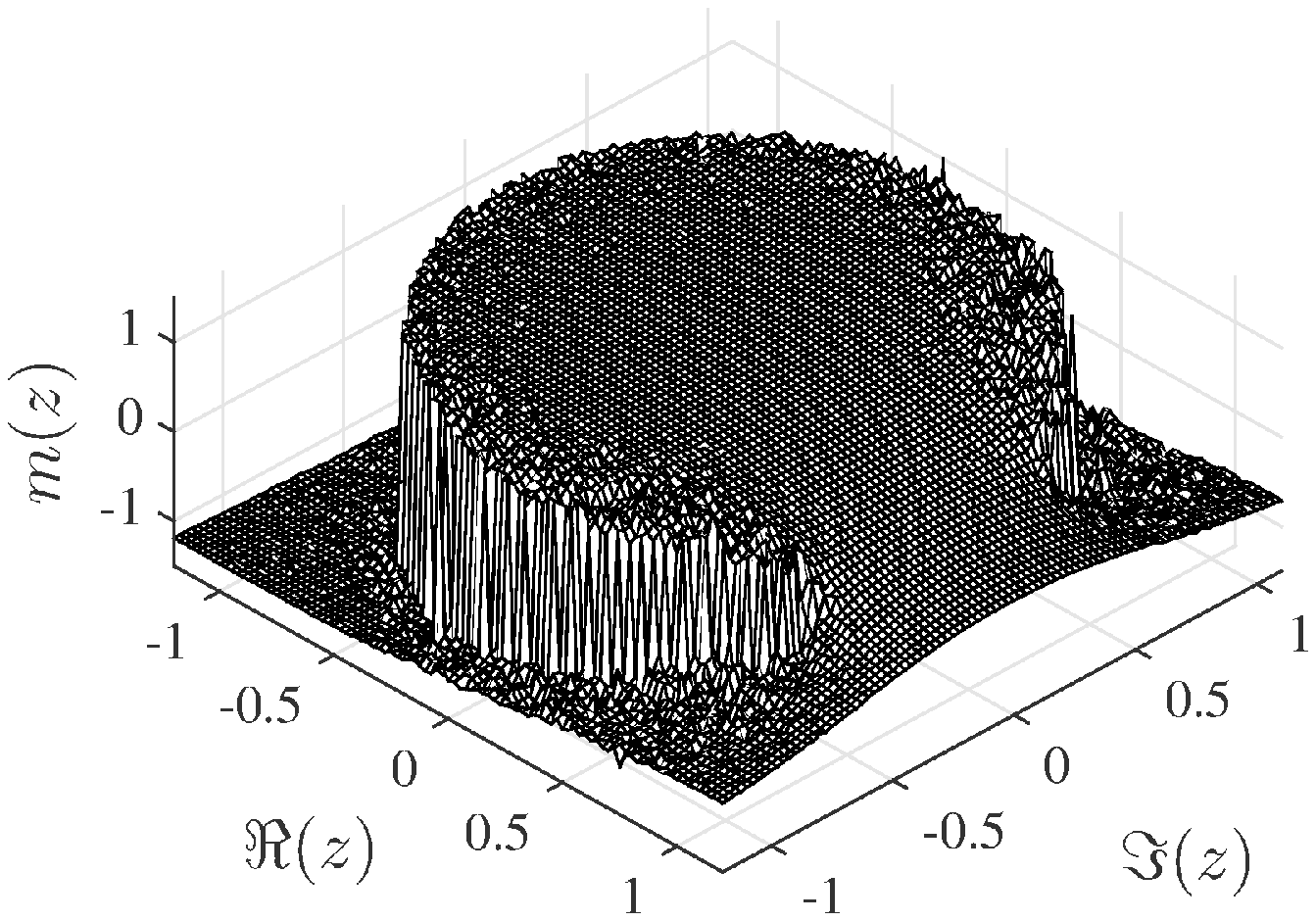}
\caption{Following Fig.~\ref{fig:mag_2D}, magnetization per site $m$ in the complex $z$ plane for the Ising model on the 3D simple 
cubic lattice. Two temperatures are fixed, $\beta=\beta_c$ (top, $D_{cut}=5$) and $\beta=\beta_c/2$ (bottom, $D_{cut}=9$).
Above the critical temperature, the discontinuity of $m$ turns into a smooth slope around
the positive real axis.}\label{fig:mag_3D}
\end{figure}

The RG method presented above can be extended to the study of higher dimensional lattices.
Using a tensor network structure resembling that of the spin lattice, the connectivity of each
component increases and the rank of the tensors becomes larger at higher lattice dimensions. 
Thus, the computational cost associated to the RG process (which depends directly on the rank and dimension of the tensors) 
is larger, and we can only reach smaller bond dimensions (and hence less precision) using the same resources, compared to 2D. 
For the Ising model in a simple 3D cubic lattice the tensors have rank $6$, and are obtained in a similar way to the 2D case, resulting in the tensor
\beq
A_{u,d,l,r,f,b} = \sum_i U_{i,u}V_{i,d}V_{i,l}U_{i,r}U_{i,f}V_{i,b}.
\eeq
Using the HOTRG procedure combining two tensors of rank $6$ at each step, the 
3D lattice gets contracted to a single tensor. The magnetization in the complex plane is obtained
also similarly as in 2D and is plotted in Fig.~\ref{fig:mag_3D}.

\begin{figure}[ht]
\includegraphics[width=9cm]{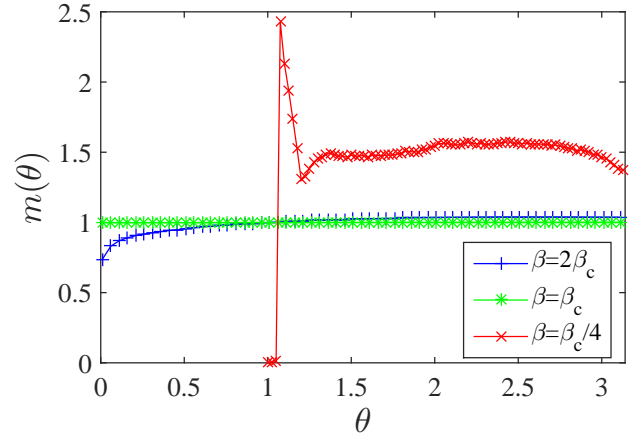}
\caption{(Color online) Magnetization $m(r\rightarrow 1^-)$ for the 2D Ising model using the TN TRG with bond dimension $D_{cut}=45$. Via the relation between the magnetzation and the density~(\ref{eqn:density2}), the main features of the density 
of Yang-Lee zeros are obtained in a  range of temperatures. For $T<T_c$, a uniform distribution of zeros is 
obtained along the unit circle. At $T=T_c$, a drop in the density near the positive real axis marks 
the presence of a phase transition. For $T>T_c$, a gap around the real axis appears and no zeros 
are found up to a critical value $\theta_e$.}\label{fig:densities}
\end{figure}

How do we then obtain the density of the zeros? The Lee-Yang theorem states that for the ferromagnetic Ising models, the partition function 
zeros lie on a unit circle in the complex $z$ plane.
The density of zeros $g(\theta)$, for any $\theta$ along the unit circle $re^{i\theta}$ with $r=1$ is related to the discontinuity of the magnetization~\cite{YangLee2,KG71}, 
\beq
\lim_{r\rightarrow 1^+}Re(m)-\lim_{r\rightarrow 1^-}Re(m)= -4\pi g(\theta). \label{eq:density}
\eeq
From our calculations we observe how the magnetization shows a discontinuity at the unit circle, 
and how this discontinuity changes with the temperature. For the Ising model, we observe that $\lim_{r\rightarrow 1^+}Re(m)=-\lim_{r\rightarrow 1^-}Re(m)$ and hence 
\begin{equation}
\label{eqn:density2}
\lim_{r\rightarrow 1^-}Re(m)= 2\pi g(\theta).
\end{equation}
Using Eq.~(\ref{eq:density}) or Eq.~(\ref{eqn:density2}) we can thus directly relate our calculation of $m$ to 
the  density of zeros along the unit circle.
Confirming the Lee-Yang theorem, our results show that the density $g(\theta)$ is zero around the real axis 
for $T>T_c$, and thus forms a gap of zero density for $\theta<\theta_e$, 
where $\theta_e$ the smallest value of $\theta$ below which no zeros occur. As the temperature increases, the zeros move towards $\theta=\pi$, and the density there keeps increasing with temperature, which is what was revealed by Eq.~(\ref{eq:z-1}). At precisely $T=T_c$ the density at the point $z=1$ 
drops to zero, indicating the temperature threshold of the existence of phase transitions. We plot in Fig.~\ref{fig:densities} the density
$g(\theta)$ for three different temperatures around $T_c$. The density of zeros is  flat below $T_c$, and exactly at
$T=T_c$ we observe a drop in the density around the real axis.  An entirely different behavior for the zeros is observed at $T>T_c$.

\begin{figure}
\includegraphics[width=9cm]{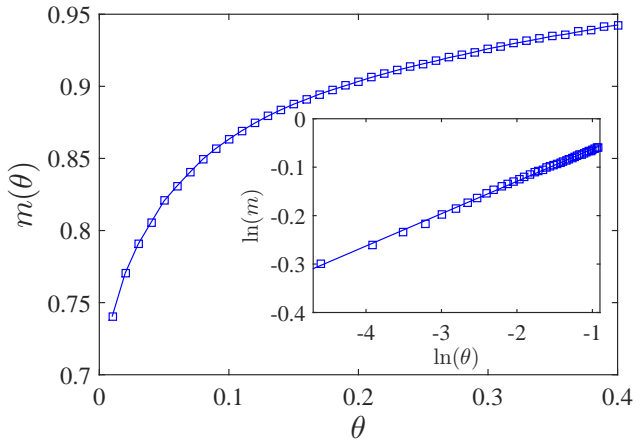}
\caption{(Color online) Magnetization along the unit circle ($r\rightarrow 1^-$) for different values of $\theta$ at exactly $\beta = \beta_c$.
Via the relation between the magnetization and the density~(\ref{eqn:density2}), the inset shows the relation $\log(m)$ vs $\log(\theta)$ following the relation $g\sim|\theta|^\frac{1}{\delta}$, with
a value of $\delta(D_{cut}=30)= 15.0(2)$ (for the Ising model in 2D, $\delta=15$).}\label{fig:2D_bc}
\end{figure}

\begin{figure}
\includegraphics[width=9cm]{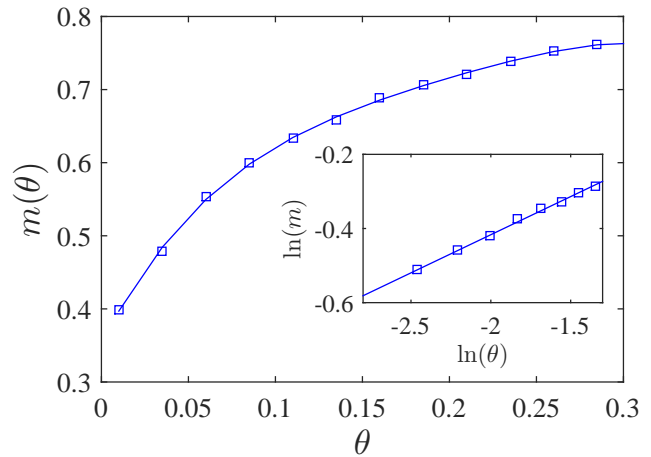}
\caption{(Color online) For an ising model in a cubic lattice,  the magnetization and hence the density of Yang-Lee zeros at the estimated critical temperature 
$T_c = 4.51154$ along the unit circle in the complex plane is plotted using a bond dimension $D_{cut}=14$. 
Via the relation between the magnetization and the density~(\ref{eqn:density2}), it is observed that this density is zero at the real axis $\theta=0$, and increases along the unit circle.
The inset shows the calculation of the critical exponent $\delta=4.8(3)$ from the relation $g\sim|\theta|^\delta$ (for a 3D Ising model in a cubic  lattice $\delta=4.789(2)$).}\label{fig:3D_bc}
\end{figure}

At the critical temperature $T=T_c$  the relation between the magnetization and the external field implies the following scaling relation of the density of the zeros,
\beq
g(\theta)\sim|\theta|^\delta
\eeq
where $\delta$ is the magnetization-field
critical exponent. In Fig.~\ref{fig:2D_bc} we plot the density at $T=T_c$, 
where we observe the detailed drop of the density at exactly $z=1$, and the inset shows the estimation of the exponent. 
From these calculations we obtain an estimation of $\delta=15.0(2)$ in agreement with the 
2D Ising magnetic exponent $\delta=15$~\cite{mccoy}. Proceeding in a similar way for the higher dimensional lattice, 
we use in our calculations an estimated critical temperature $T_c = 4.51154$~\cite{hasen,deng,gupta} for the 3D Ising model in a simple cubic lattice. 
Plotting the density at this temperature (see Fig.~\ref{fig:3D_bc}) 
we obtain the critical exponent $\delta=4.8(3)$ from the relation $g\sim|\theta|^\delta$ (for a 3D 
Ising model in a simple cubic lattice $\delta=4.789(2)$).

\section{Edge singularities at $T>T_c$}\label{sec:edge}

\begin{figure}
\includegraphics[width=9cm]{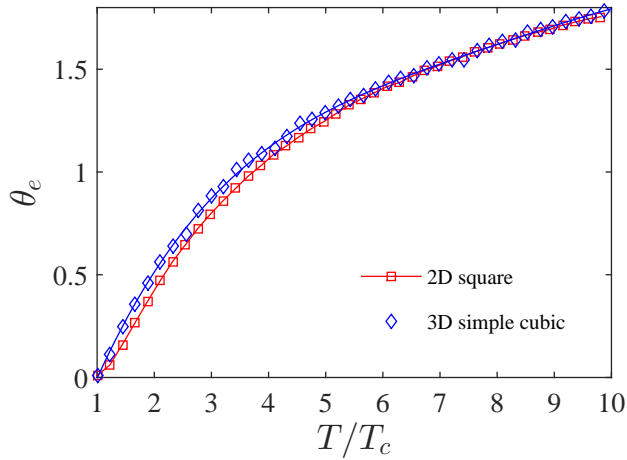}
\caption{(Color online) Location of the partition function zero closest to the positive real axis, as measured by the angle $\theta_e$.
Results for the 2D square lattice (using $D_{cut}=20$) and for a 3D simple cubic lattice (using $D_{cut}=10$) are shown, 
normalized by their respective $T_c$.
The location of the edge singularity $\theta_e$ is 
shifted towards $\theta\rightarrow \pi$ for increasing temperatures.}\label{fig:edges}
\end{figure}

The structure of the density of zeros changes dramatically at temperatures above $T_c$, 
as a gap with no zeros on the unit circle opens up around $\theta=0$, and it increases with the temperature. 
The density of the Yang-Lee zeros vanishes for $\theta<\theta_e$ and becomes non-zero starting at an edge value $\theta_e$ \cite{KG71}. 
The value $\theta_e$ can be directly determined from the density obtained using the TN method, 
and has strong dependence with the temperature. 
From the magnetization we obtain the value of $\theta_e(T)$, depicted in 
Fig.~\ref{fig:edges} for 2D square and 3D simple cubic lattices, as a function of the
temperature $T$ normalized to their respective $T_c$. 
While not exactly equal, these two curves show similar features as they both increase rapidly after $T_c$ and
then have a slower progression at larger temperatures.

\begin{figure}
\includegraphics[width=9cm]{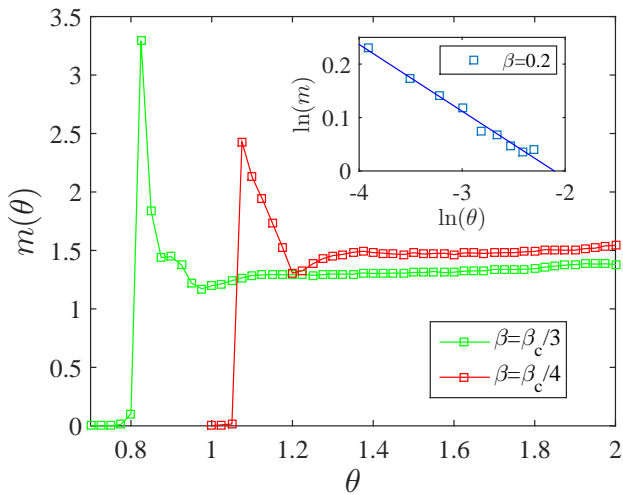}
\caption{(Color online) Density of zeros for the 2D Ising model at different temperatures above the critical value $T_c$, as computed with TN and $D_{cut}=45$. 
A gap of zeros appears around the real axis for a range of $\theta$, e.g., up to a value of $0.8$ for $\beta=\beta_c/3$. This gap extends to $\theta_e$, the edge singularity with a high density of zeros. The
value of $\theta_e$ moves towards $\theta=\pi$ as the temperature is increased.
Inset: At $\beta=0.2$, estimation of the exponent $\mu=-0.1(1)$.} \label{fig:2D_edges}
\end{figure}

In addition to $\theta_e$, there is also  a global movement of zeros away from $\theta=0$ towards the opposite side of the unit circle $\theta=\pi$ as the temperature increases. 
In Fig.~\ref{fig:2D_edges} we plot the density obtained around the unit circle as a function of $\theta$ 
for a 2D square lattice. Eventually at a sufficiently large temperature, the density will accumulate mostly at $\theta=\pi$ and diverge as $T\rightarrow \infty$. This is also verified by the
diverging behavior of $m$ at $z=-1$ in Eq.~(\ref{eq:z-1}).

The most striking feature in the behavior of the zeros is the edge singularity in 2D: as $\theta$ approaches $\theta_e$ from above, the density of zeros becomes diverging~\cite{KG71}. For increasing values of $T$, the density on the unit circle evolves as illustrated in Fig.~\ref{fig:2D_edges}. 
This singularity at $\theta_e$ has been identified as a critical point~\cite{fisher78}, and as a non-unitary realization of conformal symmetry~\cite{Car85}.
It is characterized as follows,
\beq
g(\theta) \sim (\theta-\theta_e)^\sigma, \ \ \mbox{for} \, \theta >\theta_e,
\eeq
where $\theta_e$ is the position of the divergence at each temperature. 
One expects a reduction of accuracy near a critical point, especially for this edge-singularity critical point, and we
only obtain an estimated 
value of $\sigma=-0.1(1)$, consistent with the value from conformal field considerations $\sigma=-1/6$~\cite{Car85}.  
A  different picture appears in the study for the 3D simple cubic lattice.
There is no divergence of the density and $\sigma$ is positive~\cite{KF79}; see Fig.~\ref{fig:mag_3D}b. However, due to small $D_{\rm cut}$ we can use and the noisy data from the density, we do not obtain a good estimate of $\sigma$. 

\section{Yang-Lee zeros in the Potts model with a complex field}\label{sec:potts}

In the preceding sections we have presented a method to obtain properties of the partition function zeros for lattice models, focusing on the Ising model 
in two and three dimensions. This approach can indeed be applied to any model for which we have a 
TN description, and especially in classical models where the tensors are readily obtained directly from the 
Hamiltonian expression, as explained in the Ising models above. For illustration, we shall consider the $q$-state Potts models on the square and simple cubic lattices.

The tensor network description of the $q$-state Potts model is obtained from
\beq
H = \sum_{\langle i,j\rangle} [1-\delta(\sigma_i,\sigma_j)] - h \sum_i \delta(\sigma_i,t)
\eeq
where $\sigma=0,\ldots q-1$ and $t \in [0,q-1]$ but can be chosen to be $0$ for simplicity. The transition temperature of the $q$-Potts model was found to be at $\beta_c=\log(1+\sqrt{q})$~\cite{wu}.
 Following similar considerations as in previous sections, 
we obtain a tensor network of an initial bond dimension $q$ on the square lattice (as well as the simple cubic lattice), where tensors are located at the
vertices of the lattice. The TRG method is directly applied to this tensor structure. We compute the magnetization 
\beq
m=\frac{1}{Z}{\rm Tr} \big(M\, e^{-\beta H}\big),   
\eeq
where $Z={\rm Tr} \exp(-\beta H)$ and
\beq
M=\frac{1}{N}\sum_i \delta(\sigma_i,t).
\eeq
Similar to the explanation for the magnetization in the Ising model~(\ref{eq:m}), the compuation of the magnetization in the Potts model is also a ratio between two tensor-network contractions. The density of zeros is also obtained by searching for discontinuity in the real part of the magnetization on the complex plane.

\begin{figure}
\includegraphics[width=9cm]{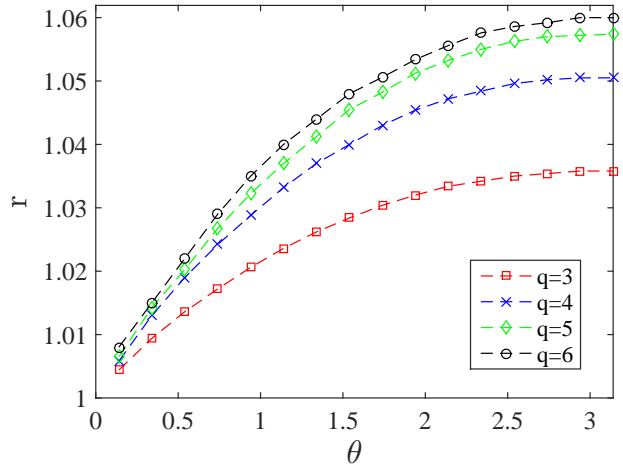}
\caption{(Color online) The location of the zeros for the 2D $q$-state Potts model using polar coordinates, as obtained with $D_{cut}=20$. At the 
corresponding critical temperature $\beta_c=\log(1+\sqrt{q})$ for each $q$, the locus of zeros lies outside the unit circle.
The location of zeros is shown for the Potts model with $q=3\ldots 6$, where the distance 
changes with $\theta$ showing a maximum value at $\theta=\pi$ for all $q$.} \label{fig:potts_circle}
\end{figure}

The exact location of Yang-Lee zeros of the Potts model has attracted attention 
as we lack the equivalence of the unit circle theorem for this model. Their location has been mostly estimated from finite-size extrapolation~\cite{KC98,Kim02}.  An interesting feature of these findings is that
for the Potts model for $q>2$  the zeros were believed to lie outside the unit circle. 
Upon increasing the temperature, the zeros move further away from the unit circle as the gap around
real positive axis opens. However, these previous results based on finite-size extrapolation were questioned~\cite{monroe}, and  calculations directly in the thermodynamic limit were not available. 
\begin{figure}
\includegraphics[width=9cm]{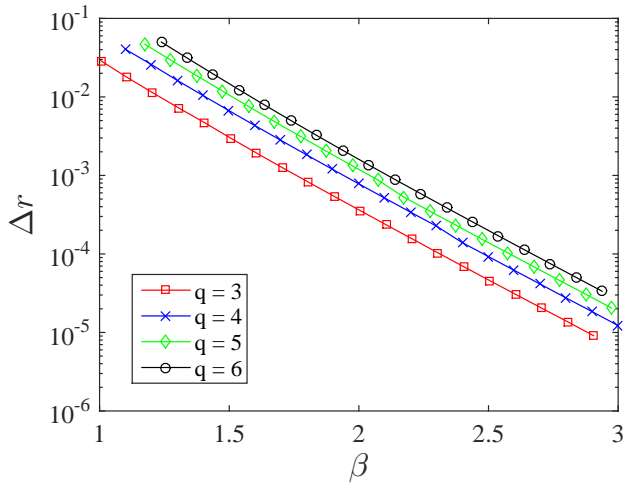}
\caption{(Color online) Distance to the unit circle $\Delta r$ in the complex plane at the point $\theta=\pi$ for the 2D $q$-state Potts model. A bond dimension $D_{cut}=20$ is used, and distances are shown for a number of temperatures $\beta$.
Increasing the value of $q$ increases the distance $\Delta r$, but decreases at higher temperatures. 
However, this plot suggests that only
in the limit $T=0$ the zeros reach the unit circle.}\label{fig:potts_distance}
\end{figure}

Our approach explained earlier enables us to probe directly the thermodynamic limit  and to resolve the debate. In Fig.~\ref{fig:potts_circle} we plot the location of zeros (obtained via the discontinuity in the magnetization) for the Potts model at $\beta=\beta_c$, for different values of $q$. 
We clearly observe how the locus of zeros is outside the unit circle, at a variable distance dependent on $\theta$ and $q$.
We have verified that our results have little dependence on the bond dimension (for $D_{cut}\gtrsim 20$), and this shows that the RG schemes, such as the HOTRG, do provide efficient method to access the thermodynamic limit. 
These results are in good agreement with those in~\cite{KC98}, providing a complete picture of  the locus of zeros at any $\theta$ in 
a  precise way. As a consequence of the large though finite correlation length at $q>4$ (of a few thousand sites~\cite{LargeCorrelation}), we use a larger bond dimension 
to achieve similar precision level for any value of $q$. As we observe in Fig.~\ref{fig:potts_circle}, the farthest zero is located at $\theta=\pi$ in agreement with previous finite-size study~\cite{KC98}. 
In Fig.~\ref{fig:potts_distance} we show the movement of the zeros located at this point as measured by $\Delta r=|r-1|$, as we lower the temperature (i.e., increase $\beta$). 
At the limit $T=0$ the zeros lie on the unit circle, but this limit is reached only exponentially slowly with the inverse temperature, at a similar rate ($\Delta r\sim e^{-4.2(2)\beta}$), for any $q$ and other 
values of $\theta$ (not shown here). 

\begin{figure}
\includegraphics[width=9cm]{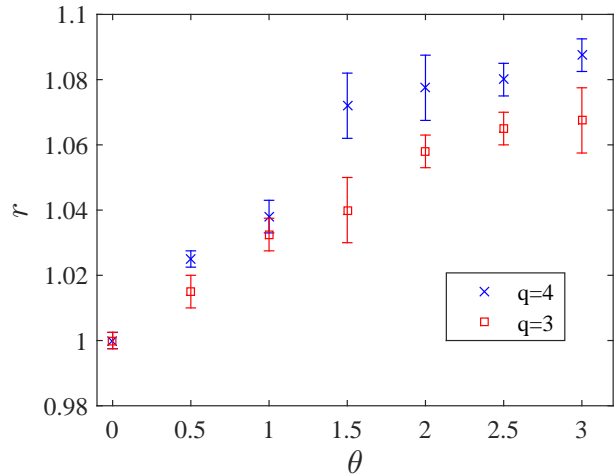}
\caption{(Color online) The location of the zeros for the $q=3,4$ Potts model using polar axis in the 3D simple cubic lattice,
as obtained up to $D_{cut}=12$. At their respective  
critical temperatures \cite{baz} the locus of zeros lie clearly outside the unit circle.}\label{fig:3D_potts_circle}
\end{figure}

In three dimensions, previous finite-size study was limited to very small system sizes such as $3\times3\times3$~\cite{Kim02}, but the zeros do not lie on the unit circle. What is their fate in the thermodynamic limit? Again we use TN methods to directly probe the zeros in this limit.  In 
Fig.~\ref{fig:3D_potts_circle} we show the locus of zeros for $q=3, 4$ Potts model in the 3D simple cubic lattice. These zeros
are computed at the critical temperature $T=T_c$ from Ref.~\cite{baz}, and we clearly see that they are located outside the unit circle in the complex $z$ plane (except at $\theta=0$). 

\section{Conclusions}\label{sec:conclusions}
We have employed the tensor renormalization methods to investigate the properties of the free energy and the distribution of the partition function zeros (i.e. the Yang-Lee zeros) in the complex field (analagous to the fugacity) plane. 
From the position and the density of Yang-Lee zeros one can determine important properties of the phase transitions of spin systems.
While the location of the zeros was rigorously established by Lee and Yang for the Ising model, their distribution has not been established precisely. We have demonstrated that the tensor network methods provide useful tools to access such information on the complex plane.

We have presented results for the density of zeros along the unit circle in the plane of complex $z=\exp\{-2\beta h\}$ in both 2D and 3D Ising models, showing  different characteristics of the density at different temperatures compared to the critical temperature.  In particular from the distribution of zeros at $T_c$ we have extracted the magnetization-field critical exponent. At higher temperatures we have determined  how the singularity edge $\theta_e$ moves with the temperature and estimated the singularity exponent. All the results are in good agreement with those from other techniques.

Going beyond the Ising models, we have also examined the $q$-state (with $q>2$) Potts model in both two and three dimensions, where fewer analytic results were known. We found that  in the thermodynamic limit the Yang-Lee zeros from these small system sizes do not lie on the unit circle except at the zero temperature, and the approach to the unit circle from high temperatures is exponentially close as the inverse temperature. This resolves a previous debate about whether the conclusion that the zeros are not on the unit circle is indeed correct in the thermodynamic limit or simply due to the finite-size effect. 
Possible future directions include the application and generalization of the approach here to other models for probing both Yang-Lee and Fisher zeros.

\smallskip\noindent {\bf Acknowledgements}.
We thank enlightening discussions with Barry McCoy and Robert Shrock. 
This work is supported by the National Science Foundation under Grant No. PHY 1314748.


\end{document}